\documentclass{aa}  

\usepackage{graphicx}
\usepackage{txfonts}
\usepackage[hidelinks]{hyperref}
\usepackage{multirow}   % Allow to do multi-row cells in tables 
\usepackage{xcolor}
\usepackage{lscape}
\newcommand{\revI}[1]{{{#1}}}
\newcommand{\revII}[1]{{{#1}}}

\begin{document}

   \title{The X-ray--UV Luminosity Relation of eROSITA Quasars}
\titlerunning{X-ray--UV relation of eROSITA QSOs}

   \authorrunning{A. Sacchi et al.}
   \author{Andrea Sacchi,
          \inst{1,2}
          Guido Risaliti,
          \inst{3,4}
          Matilde Signorini,
          \inst{5,4}
          Emanuele Nardini,
          \inst{4}
          Elisabeta Lusso,
          \inst{3,4}
          Bartolomeo Trefoloni
          \inst{4,6}
          }

   \institute{Center for Astrophysics $\vert$ Harvard \& Smithsonian, 60 Garden Street, Cambridge, MA 02138, USA
              \and
              INAF - Istituto di Astrofisica Spaziale e Fisica Cosmica Milano, Via A.Corti 12, 20133 Milano, Italy\\
              \email{andrea.sacchi@inaf.it}
              \and 
              Dipartimento di Fisica e Astronomia, Universita` di Firenze, via G. Sansone 1, I-50019 Sesto Fiorentino, Firenze, Italy \and
              INAF – Osservatorio Astrofisico di Arcetri, Largo Enrico Fermi 5, I-50125 Firenze, Italy \and
              European Space Agency (ESA), European Space Research and Technology Centre (ESTEC), Keplerlaan 1, 2201 AZ Noordwijk, the Netherlands  \and
              Scuola Normale Superiore, Piazza dei Cavalieri 7, I-56126 Pisa, Italy\\
              }

   \date{Received XX, XXXX; accepted XX, XXXX}

 \abstract{
 The non-linear relation between the UV and X-ray luminosity in quasars has been studied for decades. However, as we lack a comprehensive model able to explain it, its investigation still relies on observational efforts. 
 This work focuses on optically selected quasars detected by eROSITA. We present the properties of the sources collected in the eROSITA early data release (eFEDS) and those resulting from the first six months of the eROSITA all-sky survey (eRASS1). We focus on the subset of quasars bright enough in the optical/UV band to avoid an ``Eddington bias'' towards X-ray brighter-than-average spectral energy distributions. 
 The final samples include 1,248 and 519 sources for eFEDS and eRASS1, up to redshift $z\approx3$ and $z\approx1.5$, respectively. We found that the X-ray--UV luminosity relation shows no significant evolution with redshift, and its slope is in perfect agreement with previous compilations of quasar samples. The intrinsic dispersion of the relation is about 0.2~dex, which is small enough for possible cosmological applications. However, the limited redshift range and statistics of the current samples do not allow us to obtain significant cosmological constraints yet. We show how this is going to change with the future releases of eROSITA data. 
 }

   \maketitle

\section{Introduction}

Quasars (or quasi-stellar objects, QSOs) are extremely powerful and stable astrophysical sources known up to $z\approx10$ \citep{bogdan24}, which, in principle, would represent perfect standard candles. 

In a series of works starting from 2015 \citep{risaliti15,lusso16,lusso17,risaliti19,lusso20,bisogni21}, our group has proven that QSOs are ``standardizable'' candles through the well-known, non-linear relation between their UV and X-ray luminosity \citep{tananbaum79,lusso10}. This relation can be expressed as $L_{\rm X}\propto L_{\rm UV}^{\gamma}$ with $\gamma\approx0.6$, and it links the optical/UV emission of the disc to the X-ray emission of the corona. Coupled with a properly selected sample of quasars, this allows us to push the knowledge of the Hubble diagram beyond the $z\approx2$ limit of currently detected type Ia supernovae \citep{risaliti19}.

The method developed by our group is based on the assumption that the observed dispersion of the X-ray--UV luminosity relation is entirely due to observational issues (e.g.\ dust and gas absorption, variability, orientation of the source) and, therefore, there exists a physical mechanism that entwines the UV and X-ray emissions of AGN. Our previous work indeed shows that this assumption is well motivated, by reducing the dispersion of the relation from the historical value of 0.8 dex \citep{tananbaum79}, to $0.20-0.22$ dex by selecting only blue unobscured QSOs \citep{lusso16,lusso20}, and finally reaching \revI{0.09 dex by considering a sample with high-quality, targeted X-ray observations, and performing a full spectral analysis for each source \citep{sacchi22}}. Our group also demonstrated that the residual dispersion can be explained almost entirely by considering inclination effects and intrinsic variability of the considered sources \citep{signorini24}.

To date, however, the study and investigation of the relation rely entirely on observational efforts; in fact, despite \revI{several attempts that have been performed, a complete physical model capable of explaining the observed relation between the X-ray and UV luminosities, and describing the mechanism through which the accretion disc and the hot corona exchange energy, is still missing. Models invoking, e.g., reconnection of magnetic loops above the disc as a source of the primary X-ray radiation \citep{lusso17}, or modified viscosity prescriptions coupled with outflowing Comptonizing coronae and/or highly spinning black holes \citep{arcodia19}, seemed promising. However, they failed to simultaneously reproduce the slope, normalization, and small dispersion of the observed relation.} 

The latest compilations of our QSO samples \citep{lusso20,bisogni21} were based on the most recent releases (at that time) of the Sloan Digital Sky Survey (DR14; \citealt{paris18}), the Chandra X-ray Catalog (CXC2.0; \citealt{evans10}), and the Fourth XMM-Newton Serendipitous Source Catalog (4XMM-DR9; \citealt{webb20}), and included $\approx2500$ sources. These samples were successfully employed to build a Hubble diagram that highlighted a $\simeq4\sigma$ tension with the standard flat $\Lambda$CDM model (H$_\textup{0}=70$ km/s/Mpc, $\Omega_\textup{M}=0.3$, $\Omega_{\Lambda}=1-\Omega_{\textup{M}}$, e.g. \citealt{risaliti19,lusso20,bargiacchi21}). 

\revI{This tension was interpreted by some authors \citep{khadka20,khadka21,khadka22,montiel25} as a proof of the inapplicability of QSOs at large for cosmological studies, attributing it to either a biased selection or an evolution of the parameters of the relation with redshift. This and other forms of criticism have been addressed in depth in a dedicated publication by \citet{lusso25}, who showed that QSOs withstand as reliable standardizable candles, unless the X-ray--UV luminosity relation experiences a sudden change in its parameters at $z > 1.5$, i.e., where the extremely good agreement between quasar- and supernova-derived distances can no longer be observationally demonstrated due to the dearth of the latter probes. In other words, such a glitch would imply that QSOs below and above $z=1.5$ accrete in different ways, which is thought to be highly unlikely.} 

\revI{Irrespective of the cosmological implications, the lack of a comprehensive accretion theory behind the X-ray--UV relation demands a constant observational scrutiny of the relation itself. To this aim, in this paper, we explore the X-ray and UV properties of} a sample of QSOs detected by eROSITA (extended ROentgen Survey with an Imaging Telescope Array; \citealt{predehl21}). \revI{This effort is key, as it demonstrates both the independence of the relation from the instrument, dataset, and X-ray observational strategy (whether pointed, serendipitous, or survey mode) and the transformative potential of forthcoming eROSITA data releases.} eROSITA is the primary instrument of the Russian-German ``Spektrum-Roentgen-Gamma'' ({\em SGR}) mission, successfully launched on July 2019 \citep{sunyaev21}. With an on-axis sensitivity at 1 keV comparable with that of {\em XMM-Newton} but coupled with a wider field of view and better spectral resolution \citep{predehl21}, eROSITA represents the ideal instrument to build a vast sample of QSOs for the \revI{study of accretion physics and its possible cosmological applications}. To date, the eROSITA collaboration has published two datasets: the eROSITA Final Equatorial Depth Survey (eFEDS; \citealt{brunner22}), covering an area of 140~deg$^2$ with an average exposure time of 2~ks; and the first six months of the {\em SRG}/eROSITA all-sky survey (eRASS1; \citealt{merloni24}), whose proprietary rights lie with the German eROSITA consortium, covering half of the sky ($\approx$\,20,000~deg$^2$) with an average exposure time of 250~s.

\begin{figure}
	\includegraphics[width=\hsize]{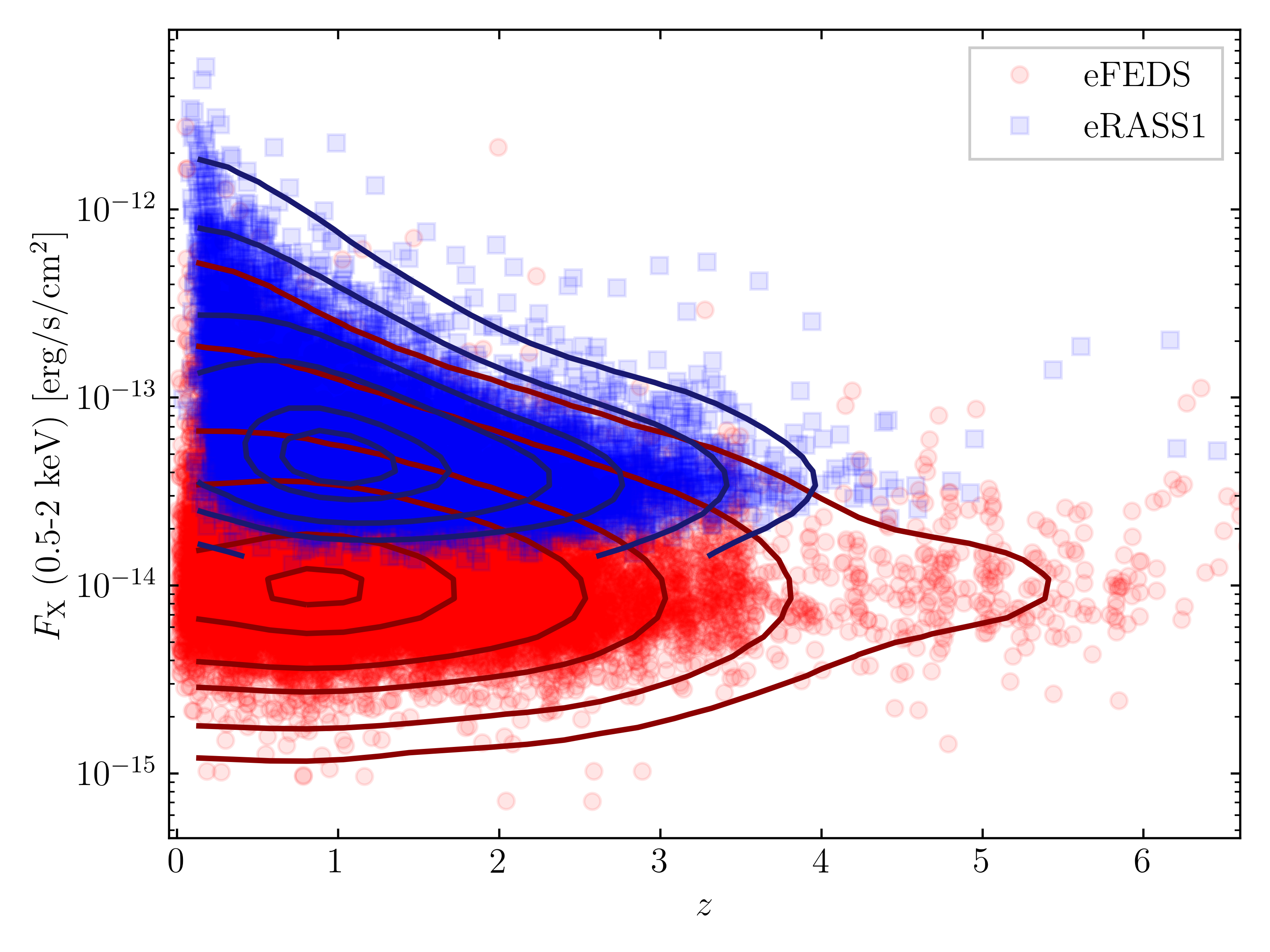}
    \caption{Flux in the 0.5--2~keV band versus redshift for the source in the eFEDS (red circles) and the eRASS1 (blue squares) parent samples. \revII{Given the high density of objects, contours with matching colors have been added to highlight the sources' distributions}.}
    \label{fig:fluxes}
\end{figure}

This paper is structured as follows. In Section \ref{sec:efeds} we report the analysis of the sources included in the eFEDS data release; in Section \ref{sec:erass1} we present the results based on eRASS1 sources; finally, in Section \ref{sec:future} we draw our conclusions and show the predictions for the future releases of eROSITA data.

\revI{Everywhere in this manuscript, the luminosity values, when reported, are computed by assuming a standard flat $\Lambda$CDM cosmology with $\Omega_\textup{M}=0.3$ and $H_0=70$~km~s$^{-1}$~Mpc$^{-1}$, unless stated otherwise.}

\section{eFEDS sources}\label{sec:efeds}

\subsection{Sample selection}

The eROSITA Final Equatorial Depth Survey (eFEDS, \citealt{brunner22}), released in 2021, covers a sky area of 140~deg$^2$ with an average exposure time of 2~ks, mimicking the expected final depth that the eROSITA all-sky survey will achieve. The full eFEDS catalog includes $\approx$\,28,000 sources; for our purposes, we employed the eROSITA Final Equatorial-Depth Survey (eFEDS) AGN Catalog \citep{liu22}, which amounts to 21,952 unique sources spanning from $z\approx0.004$ up to $z=8$ (shown in red in Fig. \ref{fig:fluxes}). 

As broadly discussed in previous works \citep{risaliti19,lusso20,bisogni21,signorini23}, we adopt the 2500~$\AA$ and 2~keV monochromatic flux densities as proxies of the accretion disc and corona emission, respectively. For the eFEDS sample, a complete X-ray spectral analysis is available, as well as the multi-wavelength coverage \citep{salvato22}. In particular, all sources possess Hyper Suprime-Cam (HSC; \citealt{aihara18}) and/or Kilo-Degree Survey (KiDS; \citealt{kuijken19}) optical photometry. Hence, we adopted the 2500~$\AA$ and 2~keV monochromatic flux densities values provided by the catalog.

\begin{figure}
	\includegraphics[width=\hsize]{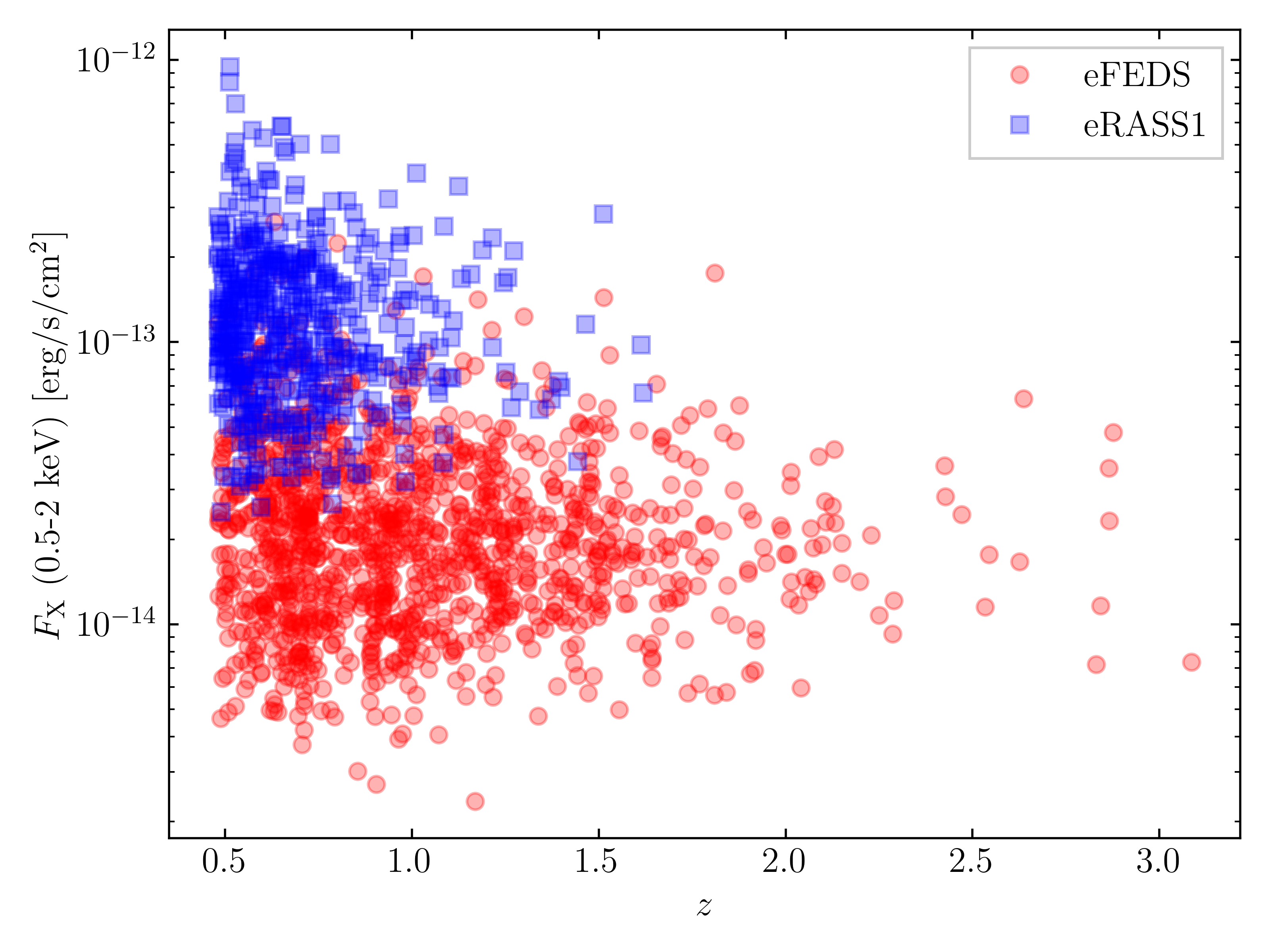}
    \caption{Flux in the 0.5--2~keV band versus redshift for the source in the eFEDS (red circles) and the eRASS1 (blue squares) final samples.}
    \label{fig:fluxes_final}
\end{figure}

The X-ray--UV relation for the sample at this stage is affected by a large scatter due to the presence of sources whose X-ray and/or UV flux is absorbed or enhanced by several different effects (see Fig. \ref{fig:rel_efeds}). Therefore, to study the intrinsic relation between disc and corona, we proceeded to clean the sample of these contaminants. To this end, we followed the approach developed in our previous works \citep{risaliti15,risaliti19} and detailed in \citet{lusso20} and \citet{bisogni21}: \revI{we removed sources affected by dust-absorption or host-galaxy contamination in the UV band, obscured in the X-ray band, or radio-loud (RL).}

\revI{The radio emission from RL QSOs can affect the estimate of the 2-keV monochromatic flux, enhancing it with respect to radio-quiet AGN \citep{zamorani81,wilkes87}. Symmetrically, objects exhibiting broad absorption lines (BALs) in the UV are thought to be more X-ray obscured with respect to non-BAL AGN (\citealt{green95,gallagher99,brandt00}; but see also \citealt{hiremath25}). In any case, the presence of BALs complicates the determination of the continuum, making the intrinsic UV emission prone to being miscalculated.} To exclude the sources affected by these effects, we cross-matched the eFEDS sample with the FIRST/NVSS catalog and the SDSS DR16Q catalog  \citep{lyke20}, and eliminated all the sources with radio-loudness parameter $R$ (=$F_\textup{6\,cm}/F_\textup{2500\,\AA}$) $>$\,10, and/or balnicity index BI\_CIV\,$>$\,0. 

\revI{As the presence of dust along the line of sight and/or the contamination by host-galaxy light could affect the measurement of the 2500-$\AA$ monochromatic flux}, we removed from the sample all the reddened sources. To this end, we computed the slope of the rest frame photometric spectral energy distribution (SED) between 1450 and 3000~$\AA$ ($\Gamma_2$). The photometric SEDs were obtained employing HSC, or, if these were not available, KIDS magnitudes. \revII{This same procedure is usually adopted to also derive the rest frame 2500~\AA\ monochromatic flux densities, but in this case this quantity is taken directly from the catalog.} We eliminated from our sample all the sources with $|\Gamma_2-0.4|>1.1$.  This choice slightly differs from the one adopted in previous works, where, along with $\Gamma_2$, also the slope of the SED computed between 0.3 and 1~$\mu$m ($\Gamma_1$) was employed to eliminate all sources outside a circle in the $\Gamma_1-\Gamma_2$ plane of radius 1.1 centered in $(0.85,0.40)$. This choice corresponded to a reddening $E(B-V)\lesssim0.1$. The first reason behind the relaxation of this requirement is that the UV/optical colour selection affects the cleaning procedure more lightly than the criteria applied to the X-ray band (such as Eddington bias and photoelectric absorption). Secondly, in doing so, we do not need infrared data in order to correctly estimate $\Gamma_1$, which largely falls in the near/mid-infrared regime starting from relatively low redshift. Thirdly, the criterion we adopted, which considers only $\Gamma_2$, still corresponds to a reddening $E(B-V)\lesssim0.1$ given that, for the SED of quasars, reddening affects more prominently the slope $\Gamma_2$ rather than $\Gamma_1$ (see \citealt{trefoloni24} for an in-depth discussion of the effects of optical/UV spectral selections and properties). To reduce possible contamination by the host galaxy, we also removed all the objects with a redshift $z\le0.48$ \citep{bisogni21}.

In all of the works presented by our group, the quality of the X-ray observations is what affects more strongly the dispersion observed in the X-ray--UV luminosity relation (see \citealt{sacchi22,signorini24} for a full discussion), therefore, the X-ray cleaning process is to be addressed with particular care. The main issue affecting the X-ray emission of QSOs is photoelectric absorption. To exclude absorbed objects, the criterion applied by our group relies on estimating the photometric slope of the X-ray spectrum ($\Gamma_\textup{X}$) and excluding all sources significantly deviating from the typical slope of unabsorbed QSOs ($\Gamma_\textup{X}$=1.9; e.g. \citealt{risaliti09}). As photoelectric absorption is naturally heavier in the soft band, flattening the X-ray spectrum and resulting in smaller values of $\Gamma_\textup{X}$, we usually excluded sources with $\Gamma_\textup{X}-\Delta\Gamma_\textup{X}\leq1.7$ \revI{(where $\Delta\Gamma_\textup{X}$ indicates the error on the photon index)}. On the other hand, too steep a spectrum could be due to observational issues, so we also exclude sources with $\Gamma_\textup{X}>2.8$. For the eFEDS sample, however, \revII{contrary to the eRASS1 sample described below (cfr. Sec. \ref{sec:erass1}), }full spectroscopic analysis is available; hence, we directly adopted the classification provided by \citet{liu22} and removed from our sample all the sources flagged as uninformative or absorbed, retaining only sources indicated as ``unabsorbed'' (\texttt{NHclass}~$=2$). In addition to this, we also excluded all objects with a low signal-to-noise ratio in the soft band (S/N~$<1$). This ensures that the X-ray spectroscopic analysis is reliable. \revII{This choice allows us to remove absorbed sources efficiently by exploiting the full potential of the complete X-ray spectroscopy available for eFEDS sources.}

\begin{table}[t!]
    \centering
    \caption{Number of sources passing each filter.}
    \begin{tabular}{l|c|c}
    \hline
     & eFEDS & eRASS1 \\ 
    \hline
    Parent sample & 21,952 & 24,375 \\
    non-BAL & 21,947 & 24,315 \\
    non-RL & 21,554 & 21,560 \\
    host contamination$^{(a)}$  & 17,657 & 18,921 \\
    X-ray selection$^{(b)}$ & 6,947 & 5,542 \\
    optical selection$^{(c)}$  & 4,859 & 5,167 \\
    Eddington bias & 1,252 & 523 \\
    \revI{$3\sigma$ clipping} & 1,248 & 519 \\
    \end{tabular}
    \tablefoot{
    \tablefoottext{a}{All sources with $z<0.48$ are excluded.}
    \tablefoottext{b}{\revII{For the eFEDS sample \texttt{NHclass}~$=2$ and S/N~$>1$ in the soft band, for the eRASS1 sample the usual photon index criterion: $\Gamma_\textup{X}>1.7$ and $\Gamma_\textup{X}<2.8$.}}
    \tablefoottext{c}{$|\Gamma_2-0.4|>1.1$ for both samples.}}
    \label{tab:filt}
\end{table}

The final factor we have to account for is the Eddington bias. QSOs exhibit intrinsic variability, and a source with an intrinsic flux close to the detection limit of the considered instrument would be detected only during a positive fluctuation of its flux. This flattens the slope of the X-ray--UV luminosity relation, as sources with low UV flux will be more easily detected in a high X-ray flux state. To correct for this bias, we employed the same strategy adopted in our previous works and described at length in \citet{lusso16}. Briefly, we assume a slope $\gamma=0.6$ for the X-ray--UV luminosity relation and adopt it to compute the expected monochromatic flux density ($f_\textup{exp}$) at 2~keV, starting from the one measured at 2500~$\AA$. Then we compare this expected flux density with the sensitivity limit ($f_\textup{lim}$) of the observation in which the object was detected. The sensitivity limit is computed based on the exposure time at the location of the source, provided by the catalog, and the sensitivity curves described in \citet{merloni12}. In practice, one should remove all the sources for which $\log f_\textup{exp}\leq\log f_\textup{lim}+k\delta$, where $\delta$ is the observed dispersion of the sample at hand, and $k$ is a multiplicative factor. \revI{The determination of the value of $k$ is clearly key in correcting for the Eddington bias, and the method through which this quantity is derived has been described at length in \citet{lusso16} and \citet{risaliti19}. As mentioned, the Eddington bias tends to flatten the slope of the relation; hence, if $k$ is underestimated, one expects a lower value of $\gamma$. On the contrary, if $k$ is overestimated, unbiased sources are removed, reducing the sample size, and increasing the intrinsic dispersion $\delta$. The optimal value of $k$ is then determined by monitoring the values of $\gamma$ and $\delta$ for different values of $k$. Once $\gamma$ remains stable and only $\delta$ increases at a further increase of $k$, the optimal value of $k$ has been reached.} 

\revI{For the eFEDS sample, we determined this optimal value to be $k=0$. This choice is equivalent to selecting sources with $f_\textup{exp}\geq f_\textup{lim}$ so, in principle, a residual bias is still present. In fact, a source experiencing a negative fluctuation is rejected, while one undergoing a positive fluctuation is retained. However, for $k=0$ the slope of the relation is already stable and compatible with the previous values obtained in the literature. This means that by increasing $k$ we obtain values of $\gamma$ that are consistent with the one obtained using $k=0$, but with reduced statistics (and therefore larger uncertainties). 
The best value being $k=0$ differs from previous studies, and can be explained given the specifics of the eROSITA sample. Contrary to samples collected by other X-ray observatories, here the flux limits have been derived using sensitivity curves published before the deployment of eROSITA, which might be slightly overestimated. Furthermore, the observational strategy of eROSITA, consisting of repeated scans of the sky, actively mitigates the variability of each individual source, making this bias less critical for the ultimate determination of the slope of the relation $\gamma$.}

The final sample \revI{(after removing four additional outliers, see below)} is composed of 1248 QSOs, shown in red in Fig. \ref{fig:fluxes_final}. A breakdown of how many sources are removed by each filter is presented in Table \ref{tab:filt}.

\subsection{X-ray--UV relation}

\begin{figure}[t]
	\includegraphics[width=\hsize]{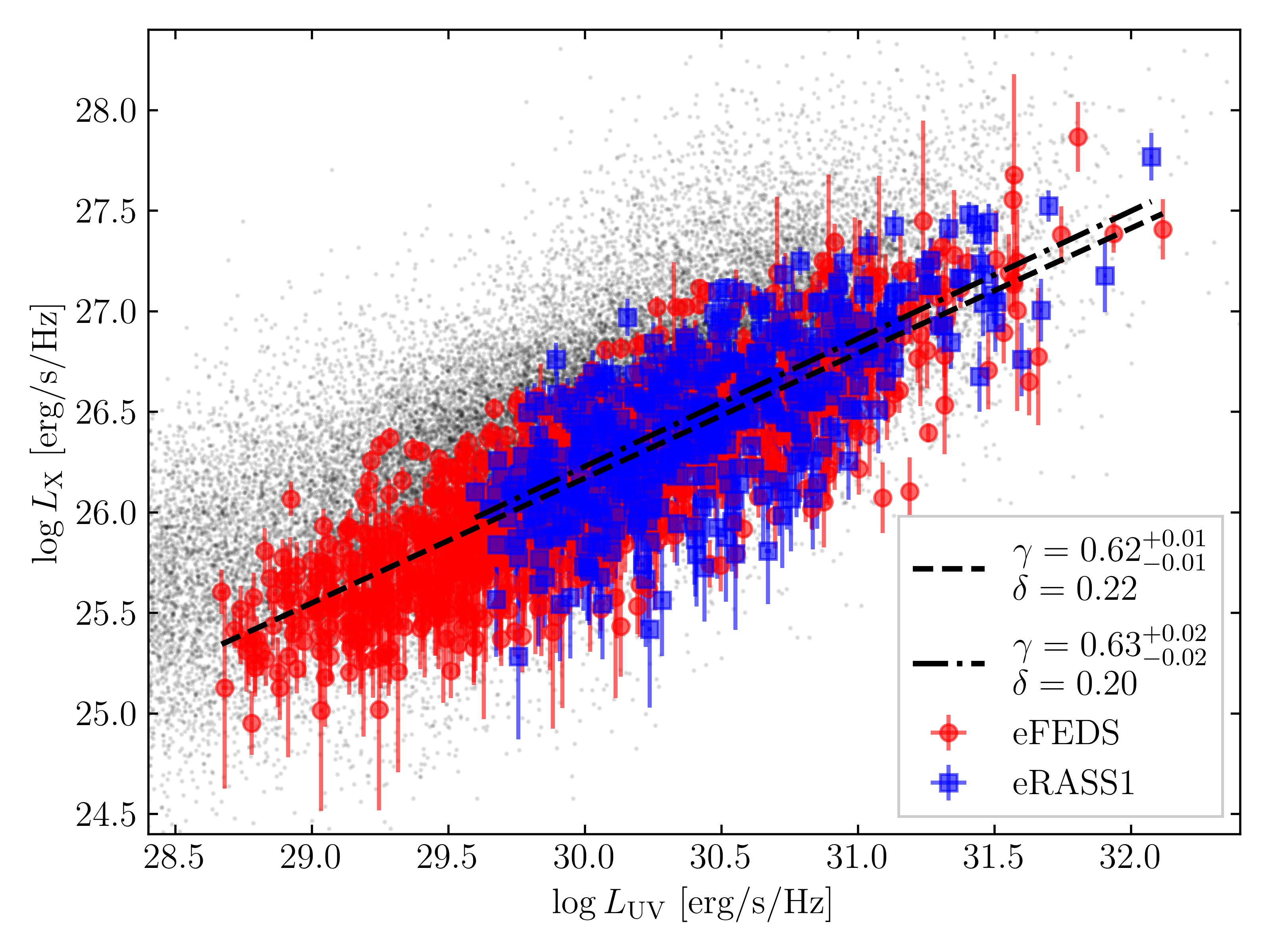}
    \caption{X-ray--UV relation. In red circles and blue squares are eFEDS and eRASS1 data, respectively. The best fits for the samples are reported by the dashed and dash-dotted lines. \revI{The gray dots represent the eRASS and eFEDS parent sample, before any filtering has been applied.}}
    \label{fig:rel_efeds}
\end{figure}

Figure \ref{fig:rel_efeds} shows the X-ray luminosity as a function of the UV luminosity (computed from the 2-keV and the 2500-$\AA$ monochromatic flux densities, respectively) for the QSOs in the eFEDS final sample (red data points). \revI{In the same Figure, in grey, we report the sources composing the two parent samples we adopted, and one can straight-away appreciate how the filtering process we adopted reduced the dispersion of the samples significantly. Furthermore, we note that the final sample is shifted towards lower values of $L_\textup{X}$. This is a further manifestation of the Eddington bias affecting the parent samples, which likely involves also the sources filtered out through the selection criteria applied to the X-ray photon index.}

To quantify this, we fitted the relation with the Python package \texttt{emcee} \citep{mcmc13}, a pure-Python implementation of Goodman \& Weare’s affine invariant Markov chain Monte Carlo (MCMC) ensemble sampler. We imposed a $3\sigma$ clipping criterion, which eliminates four outliers.  We obtained a best-fit value for the slope of the relation of $\gamma=0.62\pm0.01$. This value is in excellent agreement with those found in our group's previous works for different samples. 

Previous studies have shown that uncertainties in X-ray and UV luminosities alone cannot account for the observed scatter around the best-fit line. To model this, we introduce the parameter $\delta$, which represents the intrinsic dispersion of the relation. This dispersion arises from quasar variability, inclination effects, and the {\it intrinsic} spread of the relation itself. However, without a complete physical model, we cannot precisely quantify the latter contribution. Nonetheless, analyses of the \cite{lusso20} sample and of high-quality, smaller subsamples suggest that variability and inclination can explain a significant part of the observed dispersion, leaving little room for additional scatter from the underlying physical processes \citep{sacchi22, signorini24}.  For our sample, the best-fit value of the intrinsic dispersion is $\delta=0.22~$dex \revI{(for the combined parent sample, the dispersion is $\delta=0.36$~dex)}. This value is compatible with previous compilations of QSO samples, suggesting that the eFEDS sample is affected by the same residual biases. 

We cannot directly probe any evolution of the relation parameters with redshift; however, we can indirectly investigate possible systematic trends by splitting our sample into narrow redshift bins and employing the fluxes as proxies of the luminosities. This approach has already been exploited to prove the non-evolution of the relation up to redshift $z\approx3.5$ for the {\em XMM-Newton} and {\em Chandra} QSO samples \citep{risaliti19,lusso20,bisogni21}. We divided the sample into eight redshift bins, from $z=0.48$ up to $z=3$. The width of the bins varies to ensure that there is enough statistics in each bin and that the dispersion in distances within each bin is smaller than the one in the luminosity relation \citep[see][for a full discussion on this]{lusso25}. This guarantees that using fluxes as proxies of luminosities is a sound approximation.

\begin{figure}[t]
	\includegraphics[width=\hsize]{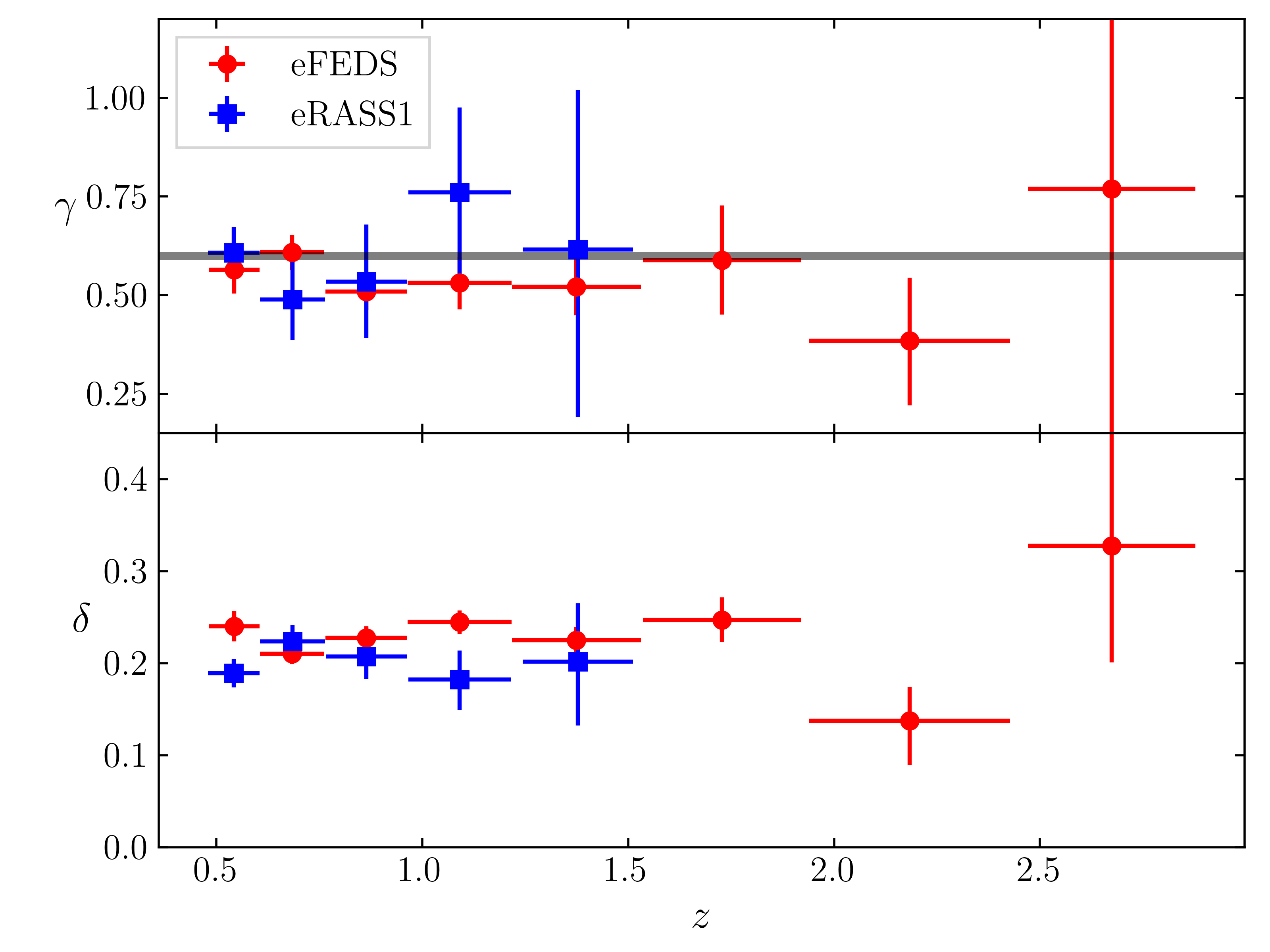}
    \caption{Parameters of the flux-flux relation as a function of redshift. The top panel shows the slope ($\gamma$), the gray line indicates the standard $0.6$ value for the slope, and the bottom panel shows the intrinsic dispersion ($\delta$). In blue and red are indicated, respectively, eFEDS and eRASS1 data.}
    \label{fig:fit_efeds}
\end{figure}

Figure \ref{fig:fit_efeds} shows, in red for the eFEDS sample, the best-fit parameters $\gamma$ and $\delta$ (slope and intrinsic dispersion) of the flux--flux relation (the fits of the objects in the individual bins are reported in Appendix \ref{app:fluxes}), along with their uncertainties. The data show no evident trend with redshift for either the slope of the relation or the dispersion. The average slope is $\langle\gamma\rangle=0.55\pm0.04$ and the average intrinsic dispersion is $\langle\delta\rangle=0.22\pm0.02$. This result confirms those obtained in previous works based on data taken with different telescopes: the slope of the relation does not evolve with redshift, and this \revI{strongly suggests that the accretion mechanism underlying the quasar phenomenon is universal. In turn, this} allows the standardization of QSOs for cosmological usage.

\section{eRASS1 sources}\label{sec:erass1}

\subsection{Sample selection}

The German eROSITA consortium released the first six months of data of the {\em SRG}/eROSITA all-sky survey (eRASS1) in early 2024. Covering half of the sky, corresponding to $\approx$\,20,000~deg$^2$ with an average exposure time of 250~s, this rich dataset includes $\approx$\,930,000 individual sources \citep{merloni24}. We built the parent sample of cosmological QSOs by cross-matching this catalog with the SDSS DR16Q catalog \citep{lyke20}, with a matching radius of 15~arcsec (corresponding to roughly half of the eROSITA point spread function in scanning mode). This produced 24,375 unique matches, represented in blue in Fig. \ref{fig:fluxes}.

We proceeded to clean this sample in the same fashion adopted for the eFEDS one. We removed BAL objects and RL sources, the former exploiting the balnicity index and the latter if they possess radio detection in the FIRST/NVSS catalog. To exclude sources that might be affected by host-galaxy contamination, all the sources with redshift $z<0.48$ are removed. 

Contrary to eFEDS objects, for the sources in eRASS1, full X-ray spectroscopic information is not available.
Hence, to remove sources affected by photoelectric absorption, we adopt the method described at length in \citet{risaliti19} and \citet{lusso20}. Briefly, we exploit the fluxes reported in the X-ray catalog in the soft (0.5--1~keV) and medium (1--2~keV) bands to derive both monochromatic 2-keV flux density \revI{(corrected for Galactic absorption)} and X-ray photon index ($\Gamma_\textup{X}$), with relative errors. To do so, the broadband fluxes are transformed into monochromatic flux densities at the respective ``pivot'' points. This ensures that the results do not depend upon the choice of photon index adopted to convert the detected count rate into energy flux, and minimizes the uncertainties on the photon index and flux. The efficiency and robustness of this procedure have been tested at length by several works from our group, and we will not address it further. We exclude all sources with $\Gamma_\textup{X}<1.7$ or $\Gamma_\textup{X}>2.8$.

Dust obscuration and Eddington bias were addressed in the same fashion as described for the eFEDS sample, \revI{while for the Eddington bias the same choice of $k=0$ was adopted}. 

Table \ref{tab:filt} presents a breakdown of the sources excluded by each filter. The final eRASS1 sample \revI{(after removing four additional outliers, see below)} includes 519 individual sources, shown in blue in Fig. \ref{fig:fluxes_final}. These are a factor of $\gtrsim$\,2 less than those in the eFEDS sample. Given that the parent samples have similar sizes, it is interesting to investigate which filters affect the size of the final samples more heavily. It is straightforward to notice that all filters remove about the same fraction of sources, except for the Eddington bias, which removes $\approx75\%$ of eFEDS sources, and $\approx90\%$ of the eRASS1 sample. In other words, 1/4th of the eFEDS sources survive the Eddington bias filter, while only 1/10th of the eRASS1 sources do. This is due to the difference in depth of the two surveys, the former being eight times deeper than the latter. 

\subsection{X-ray--UV relation}
We proceeded to fit the X-ray--UV relation for the 519 objects composing the final eRASS1 sample in the same fashion adopted for the eFEDS sample. In this case, too, we imposed a $3\sigma$ clipping cut that removed four outliers. The best-fit slope of the relation is $\gamma=0.63\pm0.02$ and the intrinsic dispersion is $\delta=0.20$. These values are both in perfect agreement with previous studies and with the eFEDS sample. Figure \ref{fig:rel_efeds} shows in blue the objects in the eRASS1 sample, and the best-fit relation is reported by a dashed-dotted line.

For this sample, too, we explored possible redshift dependencies of the relation parameters, binning the sources in narrow redshift bins in which we adopted the fluxes as proxies of the luminosities. The redshift range spanned by the sources in the eRASS1 sample is relatively narrow; hence, we can explore the properties of the relation only up to redshift $z=1.5$. Nonetheless, in the accessible range, we observed no slope evolution, and we obtained $\langle\gamma\rangle=0.58\pm0.04$ and $\langle\delta\rangle=0.20\pm0.01$, in agreement with the literature results and the average values for these parameters obtained for the eFEDS sample.\\ 

Regardless of the limited size of the presented sample (as well as of the eFEDS one), these relatively small values of intrinsic dispersion ($\delta\simeq0.2$~dex) stem from the quality of eROSITA datasets. We attribute this to the way eROSITA data are collected, performing repeated scans of the sky, de facto averaging the flux of each source. \revI{Furthermore, eROSITA has been designed for surveys, and hence the vignetting and its associated uncertainties are significantly reduced with respect to those affecting serendipitous {\it Chandra} or {\it XMM-Newton observations}}. Indeed, the values of intrinsic dispersion we obtained are comparable with previous compilations of samples of cosmological QSOs that took into account and averaged the flux of the sources over multiple observations \citep{lusso20}. This fact, coupled with the little overlap of these datasets with those obtained with other facilities (e.g., {\em XMM-Newton}, see Appendix \ref{app:xmm} for a more in-depth comparison), highlights the statistical potential of present and future eROSITA data.

\section{Future eROSITA data releases}\label{sec:future}

The analysis performed on the eFEDS and eRASS1 datasets highlights how the former outperforms the latter in terms of final sample size, depth, and redshift range. This might seem surprising, as eRASS1, although eight times shallower, covers a more than two orders of magnitude wider sky area. However, one has to consider that the luminosity function of QSOs is extremely steep, hence small increments in depth correspond to a large increase in the number of detected sources, but most importantly all of the sources in eFEDS had dedicated optical follow-up, while, for what concerns eRASS1, we had to rely on the crossmatch with the SDSS DR16Q. How much this affects our analysis can be seen by comparing the sizes of the parent samples, which, as already commented in Sec. \ref{sec:erass1}, ended up being comparable. 

While this dramatically affects the current status of publicly available eROSITA data, future releases will offer much richer opportunities. In particular, the next release of the eROSITA data will coincide with the publication of the SDSS-V \revI{(mid 2026)}, which is currently performing optical follow-ups of eROSITA-detected sources.

We can roughly estimate the improvement in the final sample size by assuming that the optical coverage will be similar, for QSOs, to the one available for eFEDS sources. As eROSITA stopped operating after completing four and a half sky scans (eRASS:4.5), we can assume an average exposure roughly equal to half of the eFEDS exposure. We can hence straightforwardly take the eFEDS clean sample and apply an Eddington-bias filter as if the exposure time was half of the actual one. This leaves 591 sources, and we can simply multiply this number by the ratio between the sky area covered by eFEDS and that covered by the full eRASS:4.5 survey, which corresponds to half of the sky. We predict that the size of a QSO sample filtered following the procedure described previously will amount to $\approx$\,78,000 sources. This is a giant leap in sample size with respect to the current state-of-the-art sample employed for cosmological studies ($\approx2000$ sources, \citealt{lusso20}), which will allow us to study the X-ray--UV relation for QSOs in unprecedented detail. 

\begin{figure}[t]
	\includegraphics[width=\hsize]{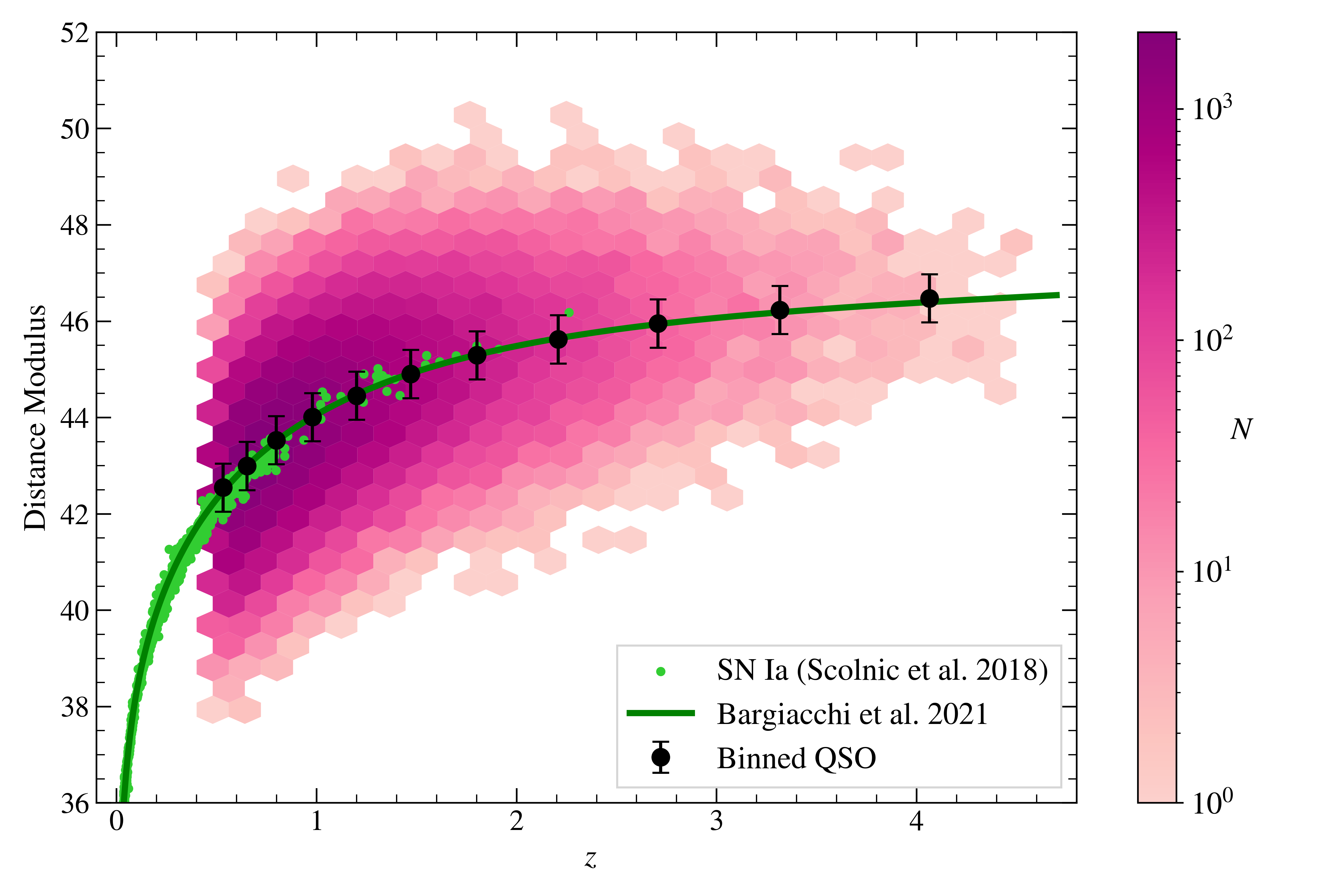}
    \caption{Hubble diagram for the simulated eRASS:4.5 sample. The density plot represents the individual sources, while averaged values are indicated by black data points. In lime are shown SNe Ia from \citet{scolnic18}. The solid green line shows the cosmographic model.}
    \label{fig:hubble}
\end{figure}

This future sample will also offer precious opportunities to compare the QSO Hubble diagram with the one built with type Ia supernovae. To grasp the potentiality of this giant improvement in sample size, we simulated the expected 78,000 sources of the eRASS:4.5 release, following the same redshift and UV flux distributions of the 591 sources in the eFEDS sample with halved exposure time. We then generated, for each simulated source, its X-ray flux \revI{assuming the cosmographic model described by \citet{bargiacchi21}}, and an X-ray--UV relation characterized by a slope $\gamma=0.6$ and a dispersion $\delta=0.2$~dex. The resulting Hubble diagram is presented in Fig. \ref{fig:hubble}. The individual sources are not reported due to their large number; instead, they are represented by a density plot. The black data points represent, in each redshift bin, the average quasar; lime points indicate the supernovae from \citet{scolnic18}, and the solid green line represents \revI{the fiducial cosmographic model}. 

\section{Conclusions}
In this work, we studied the X-ray--UV luminosity relation of unobscured, optically selected QSOs, detected by eROSITA. We compiled two different samples, based on the early data release of eROSITA observations (eFEDS) and the first data release of the mission (eRASS1). \revI{Notably, eROSITA data result from an all-sky survey, while previous compilations relied mainly on serendipitous observations.} Exploiting these samples, we confirmed the results obtained by employing independent and different observatories: the non-linear relation between X-ray and UV luminosities for unobscured, blue QSOs is tight, its dispersion entirely due to statistical and observational limits, and its parameters do not evolve with redshift. In this way, \revI{we confirmed once more that the physics of accretion is independent from redshift evolution and, as a by-product,} the exploitability of QSOs for cosmological studies, given their nature of ``standardizable'' candles. Finally, although the currently available datasets have little applicability to cosmological investigations, given the relatively limited redshift range they span, we demonstrated how the future releases of eROSITA data will represent a game changer in this domain, allowing us to build a Hubble diagram with almost 80,000 QSOs.

\begin{acknowledgements}
We thank the anonymous referee for insightful comments that improved the quality of this work. MS acknowledges support through the European Space Agency (ESA) Research Fellowship Programme in Space Science.
\end{acknowledgements}

\bibliographystyle{aa} 
\bibliography{biblio}

\appendix

\section{Comparison with {\em XMM-Newton}}\label{app:xmm}
As additional validation of the eROSITA data described in this work, we cross-match the final eFEDS and eRASS1 clean samples with the latest release of the {\em XMM-Newton} catalog of X-ray sources \citep{webb20}, obtaining 27 matches for the eFEDS sample and 30 for the eRASS1 sample, with only one source in common between the two matched samples, for which eFEDS data were considered. 

\begin{figure}[t]
 \centering
 \includegraphics[width=\hsize]{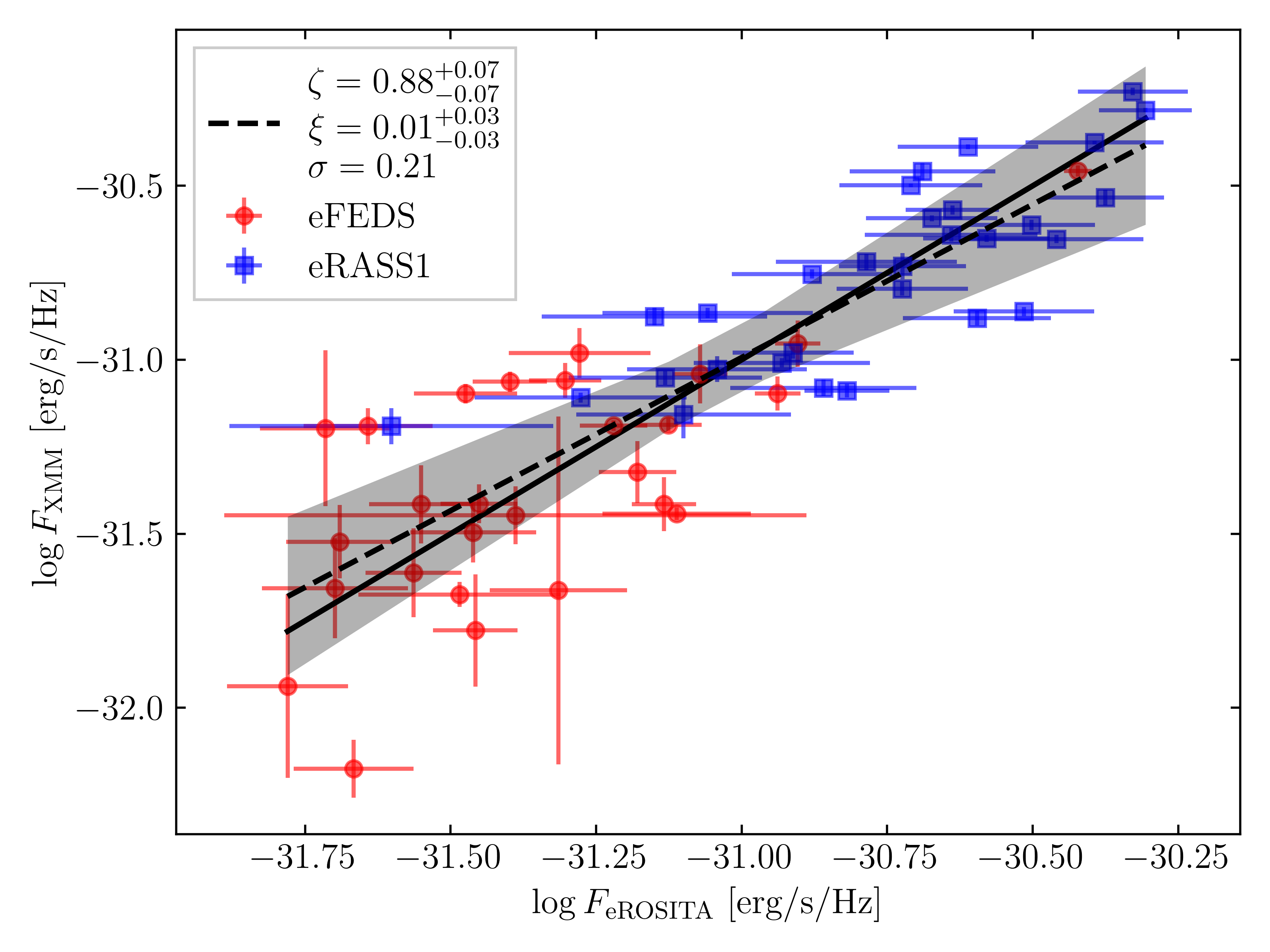}
 \caption{Monochromatic 2~keV flux densities obtained from {\em XMM-Newton} versus eROSITA data. The solid black line indicated the bisectrix. The best fit is shown by the dashed line, and $3\sigma$ uncertainties are indicated by the shaded area. Red circles and blue squares indicate whether eROSITA data come from eFEDS or eRASS1, respectively. The legend also reports the values obtained for the slope ($\zeta$) and intercept ($\xi$).}
 \label{fig:xmm}
\end{figure}

Figure \ref{fig:xmm} shows the 2-keV monochromatic flux densities obtained from {\em XMM-Newton} and eROSITA, respectively. For {\em XMM-Newton} sources, the flux was computed by interpolating between soft and hard energy bands as per usual. The fluxes obtained with the two independent telescopes are in excellent agreement, and the best fit, indicated by the dashed line in the figure, is compatible with the bisectrix (indicated by the solid line). 

A comparison with a similar analysis, shown in Fig. 13 of \citet{merloni24}, highlights that for the sources presented in this work, the ratio between eROSITA and {\em XMM-Newton} fluxes is much closer to unity. In particular, in \citet{merloni24}, the sources with lower flux levels have, on average, significantly higher fluxes measured by eROSITA with respect to {\em XMM-Newton}. This is because {\em XMM-Newton} observations are, on average, much deeper than eROSITA's ones, hence the detections of the latter are strongly affected by the Eddington bias. In compiling our cosmological sample of QSOs, we cleaned our sample from all sources affected by the Eddington bias, hence obtaining a much better compatibility between eROSITA and {\em XMM-Newton} fluxes. \revI{Our analysis, spanning almost two orders of magnitude in flux}, highlights the excellence of eROSITA's detectors\revI{: any systematic effect, for example due to detector defects, can be excluded given that the intercept $\xi$ of the relation between {\em XMM-Newton} and eROSITA fluxes is compatible with 0. All the flux discrepancies are entirely due to statistical fluctuations and} intrinsic AGN variability, given that {\em XMM-Newton} observations predate eROSITA's by several years.

\section{Flux--flux relation}\label{app:fluxes}
\begin{figure*}[t]
 \centering
 \includegraphics[width=\textwidth]{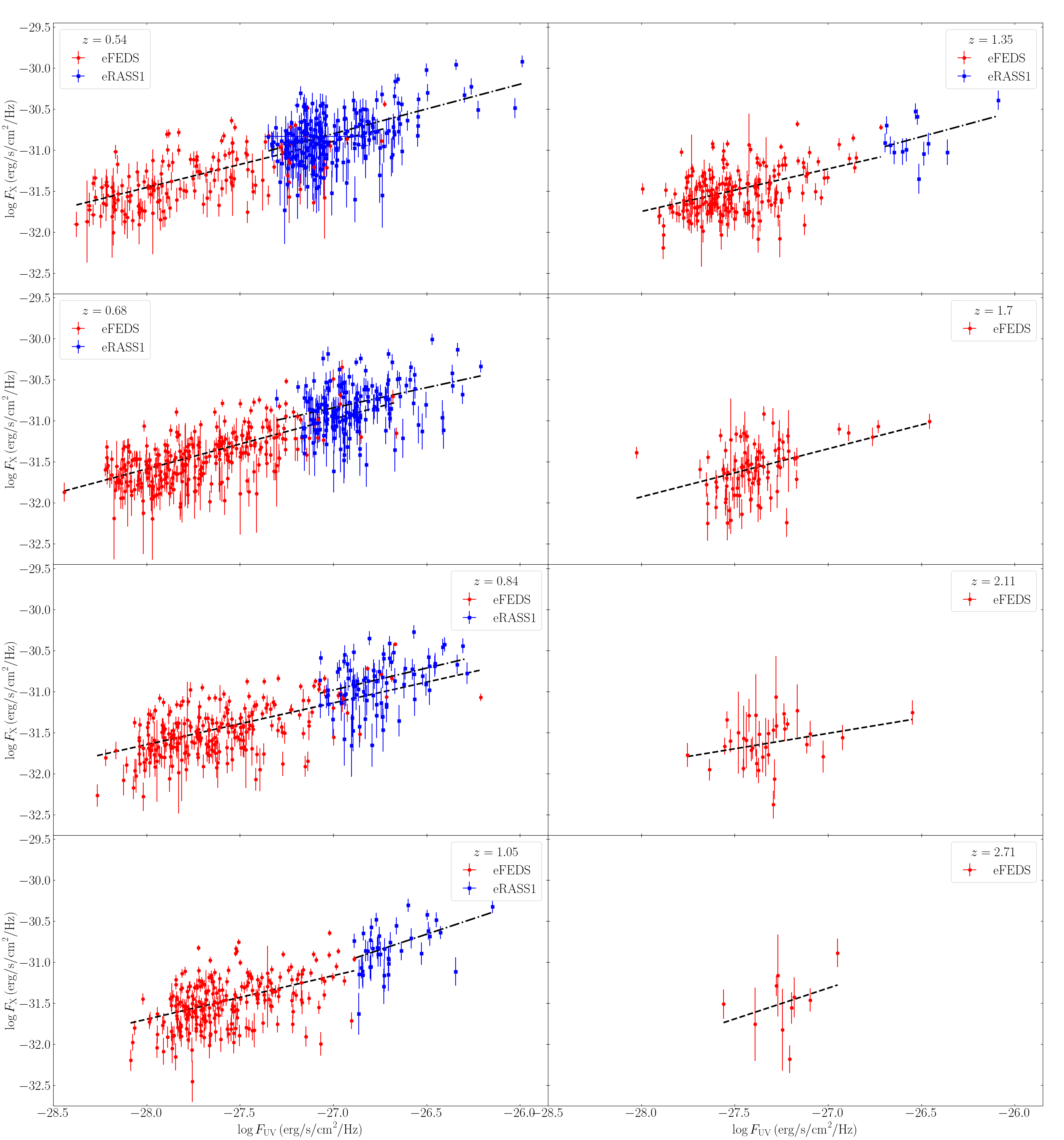}
 \caption{Fit to the X-ray--UV fluxes relation for the eFEDS (red circles) and eRASS1 (blue squares) samples in narrow redshift bins. The best fits are reported by dashed and dash-dotted lines, respectively.}
 \label{fig:flux_rel}
\end{figure*}

Figure \ref{fig:flux_rel} shows the fit of the flux--flux relation in narrow redshift bins. The best-fitting parameters of the relations are reported in Fig. \ref{fig:fit_efeds}.

\end{document}